\documentclass[aps,preprint,showpacs,superscriptaddress,showkeys]{revtex4}

\usepackage{verbatim}
\usepackage{epsfig}
\usepackage{amssymb}
\usepackage{amsmath}
\usepackage{subfigure}

\newcommand{\bea}{\begin{eqnarray}}
\newcommand{\eea}{\end{eqnarray}}

\begin{document}

\title{Detecting the traders' strategies in Minority-Majority games
and real stock-prices.}

\author{V.~Alfi}
\email[E-mail address: ]{Valentina.Alfi@Roma1.infn.it}
\affiliation{Universit\`a ``La Sapienza'', Dip. di Fisica,  P.le A. Moro 5,
  00185, Rome, Italy.}
\affiliation{Universit\`a ``Roma Tre'', Dip. di Fisica,  V. della Vasca Navale
  84, 00146, Rome, Italy.}
\author{A. De Martino}
\affiliation{Istituto dei Sistemi Complessi - CNR, via dei Taurini 19,  00185, Roma, Italy.}
\author{L. Pietronero}
\affiliation{Universit\`a ``La Sapienza'', Dip. di Fisica,  P.le A. Moro 5,
  00185, Rome, Italy.}
\affiliation{Istituto dei Sistemi Complessi - CNR, via dei Taurini 19,  00185, Roma, Italy.}
\author{A.~Tedeschi}
\affiliation{Istituto dei Sistemi Complessi - CNR, via dei Taurini 19,  00185, Roma, Italy.}

\begin{abstract}
Price dynamics is analyzed in terms of a model which includes the possibility
of effective forces due to trend followers or trend adverse strategies.
The method is tested on the data of a minority-majority model and indeed
it is capable of reconstructing the prevailing traders' strategies in a given
time interval.
Then we also  analyze real (NYSE) stock-prices dynamics and it 
is possible to derive an indication for the 
the ``sentiment'' of the market for time intervals of at least one day.
\end{abstract}

\pacs{89.75.-k, 89.65+Gh, 89.65.-s, 05.20.-y}
\keywords{Complex systems, Time series analysis, Financial data, Economic systems}
\maketitle

\section{Introduction}

The simplest representation of price dynamics is usually
considered as a simple Random Walk (RW).
It is easy to realize, however, that many important deviations
ara also present.
The most studied are the problem of the ``fat tails'' (in the
distribution of price returns), the volatility clustering and
various other elements related to the non stationarity
of the process~\cite{mantegna-stanley,bouchaud-potters}.
The arbitrage condition implies that no simple correlations can be present.
A large effort is therefore devoted to the
identification of complex correlations of various types.
These correlations arise from the collective behavior of traders, which 
lastly, define the price.

In this perspective a simple classification of trading
strategies can be made in terms of trend followers
or trend adverse.
Usually these different strategies are taken as input in models
which represent the behavior of traders.

Here we would like to consider the complementary point of view.
Namely, given a time series, is it possible to
identify, from the data, the strategies of the traders?
In order to address this question we use a new approach which is based 
on a RW plus a force which depends on the 
distance of the price from some suitable
moving average~\cite{vale1,taka1}.
This idea is that, with such an analysis, one can identify
the ``sentiment'' of the market in a given time interval.

In this paper, we first perform some statistical tests of
the method to  clear its signal to noise ratio.
Then we apply the method to time series generated by a minority-majority
model~\cite{ameno}. 
This is an important test because, in this case, one
knows the prevailing strategy of the traders.
The results are rather encouraging because the method can indeed identify
these strategies.
Finally we apply the method to real
stock prices data of the NYSE and the preliminary results show that
it is possible to derive statistically significant information
on the prevailing trading strategy for a single day or larger
time periods.

\section{The Effective Potential Model}

In recent papers~\cite{taka1,vale1} has been introduced
the idea the the stock-price dynamics can be influenced
by a moving average of the price itself in the previous time steps.
Hence, at every time step $t$, one can introduce a moving
average of the previous $M$ time steps:

\bea
{P_M(t)}=\frac 1M \sum_{\tau=0}^{M-1}{P(t-\tau)}
\label{eq:mmobile}
\eea

\begin{figure}
  \begin{center}
    \resizebox{80mm}{!}{\includegraphics{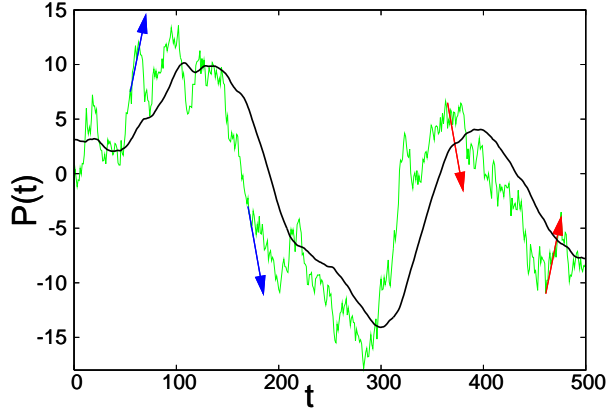}}
    \caption{Example of a model of price dynamics (in this case a simple 
      RW) together with its moving average defined as the average
      over the previous $50$ points. The idea is that the distance of the
price from its moving average can lead to repulsive (blue arrows) or
      attractive (red arrows) effective forces.}
    \label{fig:mmobile}
  \end{center}
\end{figure}

In Fig. ~\ref{fig:mmobile} is plotted the time evolution
of a RW with Gaussian random noise together
with the moving average of the price.

One can investigate if there could be a relation between the
next price increment $P(t+1)-P(t)$ and the difference 
$P(t)-P_M(t)$.

The simplest assumption is to adopt a linear dependence:

\bea
P(t+1)-P(t)\propto P(t)-P_M(t)
\label{hypo}
\eea

In this case, the price dynamics can be described
in terms of a RW with the existence of a linear
force. This force can be either repulsive or attractive 
depending on the sign of the constant of proportionality
between $P(t+1)-P(t)$ and $P(t)-P_M(t)$.

Therefore, the dynamical equations of the price is 
a RW with the presence of a force that is
the gradient of a quadratic potential $\Phi$~\cite{taka1}.

\bea
P(t+1)=P(t)-b(t)\frac{d}{dP(t)}\Phi \Big(P(t)-P_M(t)\Big)+\omega(t)
\label{eq:stoch}
\eea

where $\omega(t)$ corresponds to a random noise with unitary variance
and zero mean. $P_M(t)$ is the moving average described in 
Eq.~\ref{eq:mmobile}.

The potential $\Phi$ together with the pre-factor $b(t)$ describe
the interaction between the price and the moving average. In simple
assumption  of a linear force~\cite{taka1},
$\Phi$ results to be quadratic:
\bea
\phi\Big(P(t)-P_M(t)\Big)={\Big(P(t)-P_M(t)\Big)}^2.
\label{eq:potential}
\eea

We can simulate a process whose dynamical stochastic
equation is given by Eq. \ref{eq:stoch}.

\begin{figure}
  \begin{center}
    \resizebox{80mm}{!}{\includegraphics{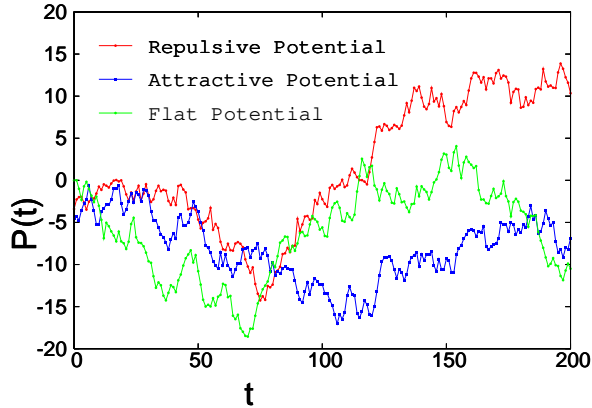}}
    \caption{Time evolution of the price described by Eq.~\ref{eq:stoch}.
Three different behavior are plotted. The red lines represents
the time evolution of a RW in a repulsive quadratic potential while
the blue line is in an attractive quadratic potential. The green line
is the case of flat potential (simple RW).
The parameters are fixed to $M=20$ and $b=\pm 1$.
We can observe an over diffusion (under diffusion) in the case of repulsive
(attractive) potential.}
    \label{fig:plot}
  \end{center}
\end{figure}

The time evolution of the ``price'' of such a process, is shown
in Fig. \ref{fig:plot}, where we can observe three cases in which
the potential is attractive, repulsive and constant (simple RW).

From this simulation, one can reconstruct the force of
the process plotting $P(t+1)-P(t)$ as a function of $P(t)-P_M(t)$.
Then, integrating from the center, one can obtain the potential.

\begin{figure}
  \begin{center}
    \resizebox{80mm}{!}{\includegraphics{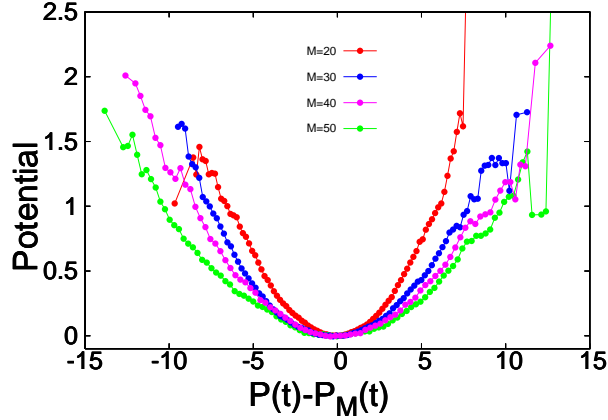}}
    \caption{The plot shows the shapes of the quadratic attractive potentials
defined by Eqs.~\ref{eq:stoch} and ~\ref{eq:potential}. We can see that
the amplitude of the potentials depend on the choice of the parameter $M$.}
    \label{fig:pote}
  \end{center}
\end{figure}

\begin{figure}
  \begin{center}
    \resizebox{80mm}{!}{\includegraphics{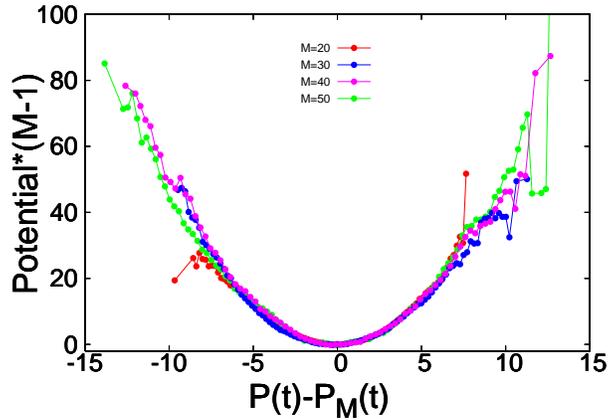}}
    \caption{The different potential plotted in Fig.~\ref{fig:pote} are 
re-plotted scaling the potential with the factor $(M-1)$. We can
see that in this way we obtain  a good data collapse.}
    \label{fig:poteM}
  \end{center}
\end{figure}

In Fig. \ref{fig:poteM} are shown the potential obtained
from a simulation of the process described in Eq. \ref{eq:stoch}
in the case of attractive force for various values of $M$.
We can observe that the potentials have an amplitude
(that is the slope of the linear force) which depends
on $M$.
In \cite{taka1} is shown that such a dependence can be 
eliminated rescaling the potential by a factor $(M-1)$.

In Fig.~\ref{fig:poteM} are shown the potentials plotted
in Fig.~\ref{fig:pote}, rescaled by the factor $(M-1)$.
Indeed we can observe a good data collapse.

This idea of assuming a linear force in Eq.~\ref{eq:stoch}  has been tested
on real data.
In ~\cite{taka1} a series of data from the Yen-Dollar exchange rates
have been analyzed.
The potential analysis for the case of the Yen-Dollar exchange rates indeed
leads to the observation of rather quadratic potentials.

Anyhow, other kind of force can be considered.
For example one can suppose that the price dynamic could depend
only on the sign of the difference $P(t)-P_M(t)$.
In this case an interesting model is represented by the following
dynamic for a RW with only up and down steps~\cite{vale1}.

\bea
\Bigg\{
\begin{array}{ll}
p(\uparrow)=1/2+\epsilon_1\;\;\;\mbox{for}\;\;\;P(t)-P_M(t)>0\\
p(\downarrow)=1/2-\epsilon_1
\label{eq:sign1}
\end{array}\\
\Bigg\{
\begin{array}{ll}
p(\uparrow)=1/2-\epsilon_2\;\;\;\mbox{for}\;\;\;P(t)-P_M(t)<0\\
p(\downarrow)=1/2+\epsilon_2\;.
\label{eq:sign2}
\end{array}
\eea

This model implies a tendency of destabilization
(or stabilization) depending on the signs of $\epsilon_1$ and 
$\epsilon_2$.

\begin{figure}[h]
\centering
\subfigure[]{
\label{fig:signA}
\includegraphics[width=3in]{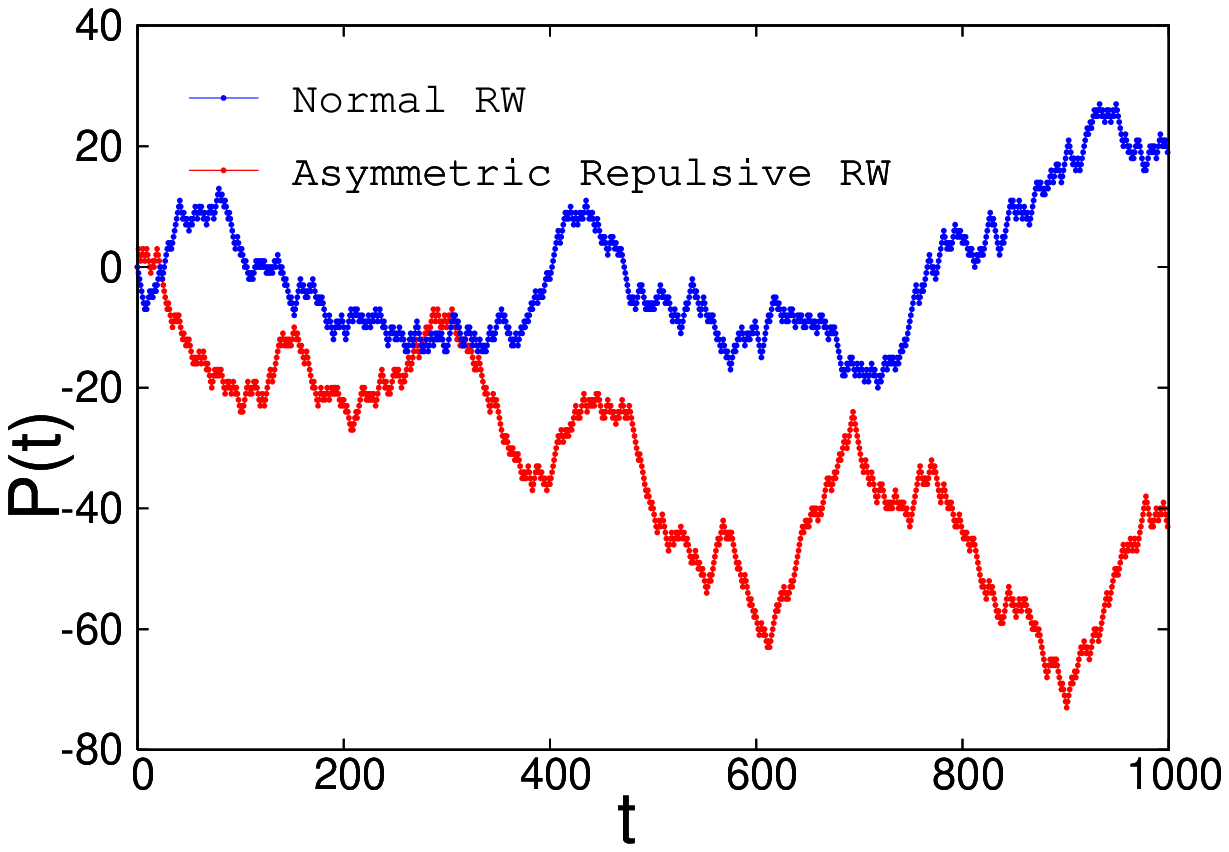}}
\subfigure[]{
\label{fig:signB}
\includegraphics[width=3in]{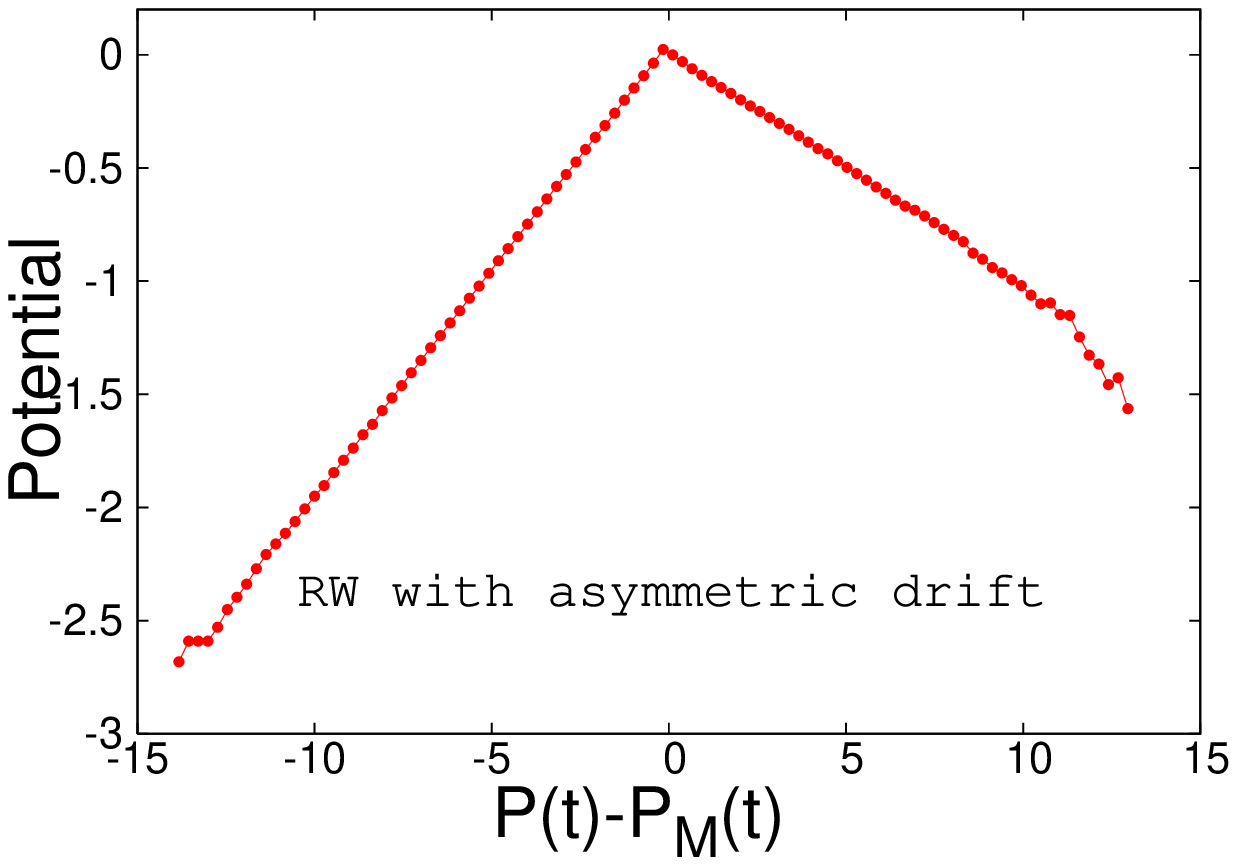}}
\caption{In figure (a) is plotted the time evolution the price whose dynamics
is described by Eqs.~\ref{eq:sign1} and ~\ref{eq:sign2}, compared
with the time evolution of a simple RW. In figure (b) is plotted the
shape of the potential obtained analyzing the data plotted in (a). We
can see a piecewise linear shape because the relative model
depends only on the sign of the difference $P(t)-P_M(t)$.
The slope of the two lines of the potential depends on the 
parameters $\epsilon_1$ and $\epsilon_2$.}
\label{fig:sign}
\end{figure}

The potential analysis for this case leads to  a piecewise
linear potential in which the slopes are related to $\epsilon_1$ and 
$\epsilon_2$. The potential will be asymmetric if $\epsilon_1\neq \epsilon_2$.
In Fig.~\ref{fig:signA} and  ~\ref{fig:signB} are shown
the time evolutions of a price whose dynamical equations is
given by Eqs.~\ref{eq:sign1} and ~\ref{eq:sign2} in the
case of asymmetric repulsive potential, and the relative
shape of the potential.

In order to test this model, we have considered an agent based model
and we have performed the potential dynamics on the time
series of the theoretical price that comes from 
the simulations.
In the next section we are going to describe the specific rules
of the agent based model we have considered.

\section{Application and test on an agent based model}

It is instructive to analyze the effective potential scenario in
agent-based models, where the price process is not defined explicitly
but only through the aggregate choices of a group of traders. The
simplest and most studied framework from a statistical physics
perspective is that of Minority Games \cite{rev,book1}, in which each
of $N$ agents must decide at every (discrete) time step whether to buy
($a_i(t)=1$) or sell ($a_i(t)=-1$) an asset. The resulting price
process is determined by the decisions of all agents through the
``excess demand'' $A(t)=\sum_{i=1}^N a_i(t)$. In particular,
neglecting liquidity effects for the sake of simplicity, one can write
that

\begin{equation}
P(t+1)-P(t)=A(t),
\end{equation}

which amounts to defining the (log-)price as
$P(t)=\sum_{t'<t}A(t')$. 

It is clear that an agent's trading behavior
will depend on his expectations about the future price increment
$A(t)$, denoted by $\mathbb{E}_i[A(t)]$. For example,
it has been argued \cite{matteopa} that if

\begin{equation}
\label{sock}
\mathbb{E}_i[A(t)]=\psi_i A(t-1)+(1-\psi_i)A(t-2),
\end{equation}

agent $i$ behaves as a trend-follower for $\psi_i>1$ (correspondingly 
he perceives the market as 
a Majority Game with payoff $\pi_i(t)=a_i(t)A(t)$), while he behaves
as a fundamentalist for $0<\psi_i<1$ and plays a Minority Game
with payoff $\pi_i(t)=-a_i(t)A(t)$. 

Let us now consider an agent who forms expectations according to
(\ref{hypo}). It is easy to see that such an agent is described by a
generalization of (\ref{sock}). Indeed, a direct calculation shows
that (\ref{hypo}) corresponds to 

\bea
\label{A_rel}
\mathbb{E}_i[A(t)]\propto\sum_{\tau=1}^{M-1}\frac{M-\tau}{M} A(t-\tau).
\eea 

Agents thus tend to discount events further back in
time and give larger weight to recent price changes when estimating
the future returns. Clearly, such an agent has a more complicated
reaction pattern than a pure Minority or Majority Game player and will
be described by a payoff function that accounts for the possibility of
behaving differently in different market regimes.

Models of this type have been introduced recently and appear to be an
ideal testing ground to verify the emergence of the effective potential
scenario in a microscopic setting. Specifically, we have tested it on
a model in which agents may switch from a trend-following to a
fundamentalist attitude (and vice-versa) depending on the market
conditions they perceive, which was introduced in
Ref. \cite{ameno}. We refer the reader to the literature for a
detailed account of the model's definition and properties. In a
nutshell, it describes agents who strive to maximize the payoff

\begin{equation}
\pi_i(t)=a_i(t)[\mathcal{A}(t)-\epsilon\mathcal{A}(t)^3],
\end{equation} 

where $\mathcal{A}(t)=A(t)/\sqrt{N}$ is the normalized
excess demand. 
The idea is that for small price movements ($A(t)\simeq
0$) agents perceive the game as a Majority Game as they try to
identify profitable trends. However when price movements become too
large, the game is perceived as a Minority Game, i.e. agents expect
the price to revert to its fundamental value. As in most Minority
Games, agents have fixed schemes (`strategies') to react to the
receipt of one of $P$ possible external information patterns and learn
from experience to select the strategy and, in turn, the action
$a_i(t)$ that is more likely to deliver a positive payoff. A realistic
dynamical phenomenology is obtained in a whole range of values of the
model parameter $\epsilon$ when the number $N$ of players is large
compared to the amount of information available to them $P$ (this is
measured by a parameter $\alpha=P/N$, see \cite{ameno} for details).

\begin{figure}
  \begin{center}
    \resizebox{80mm}{!}{\includegraphics{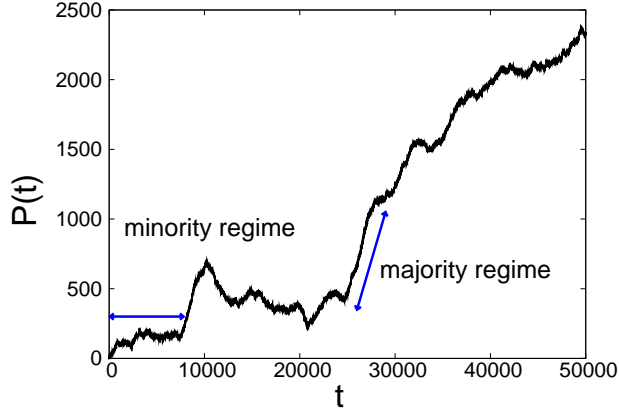}}
    \caption{The time evolution of $P(t)$ for a minority-majority
game with $\epsilon=1$ and $\alpha=0.05$ is shown. We can observe 
the alternation of different regimes. In the graph are indicated
two periods by means of arrows. In the minority regime the price remains
near to its fundamental value, while in the majority regime appears
a well defined trend.}
    \label{fig:grafminmaj}
  \end{center}
\end{figure}

In Fig.~\ref{fig:grafminmaj} is shown the time evolution of $P(t)$ for a
game with parameters $\alpha=0.05$ and $\epsilon=1$.
This choice of parameters corresponds to be in the range
in which the competition between trend followers and contrarians
is stronger.
In fact, in Fig.~\ref{fig:grafminmaj}, we can observe some ``ordered''  periods,
where  $A(t)$ is small and well defined trends in the price dynamics 
appear, but also ``chaotic'' periods where the dynamics
of the price is dominated by the contrarians.
In Fig.~\ref{fig:grafminmaj} we have identified two periods in which
the different behaviors of the agents are well defined and we have used
these periods as dataset for our potential analysis.

\begin{figure}[h]
\centering
\subfigure[]{
\label{fig:min}
\includegraphics[width=3in]{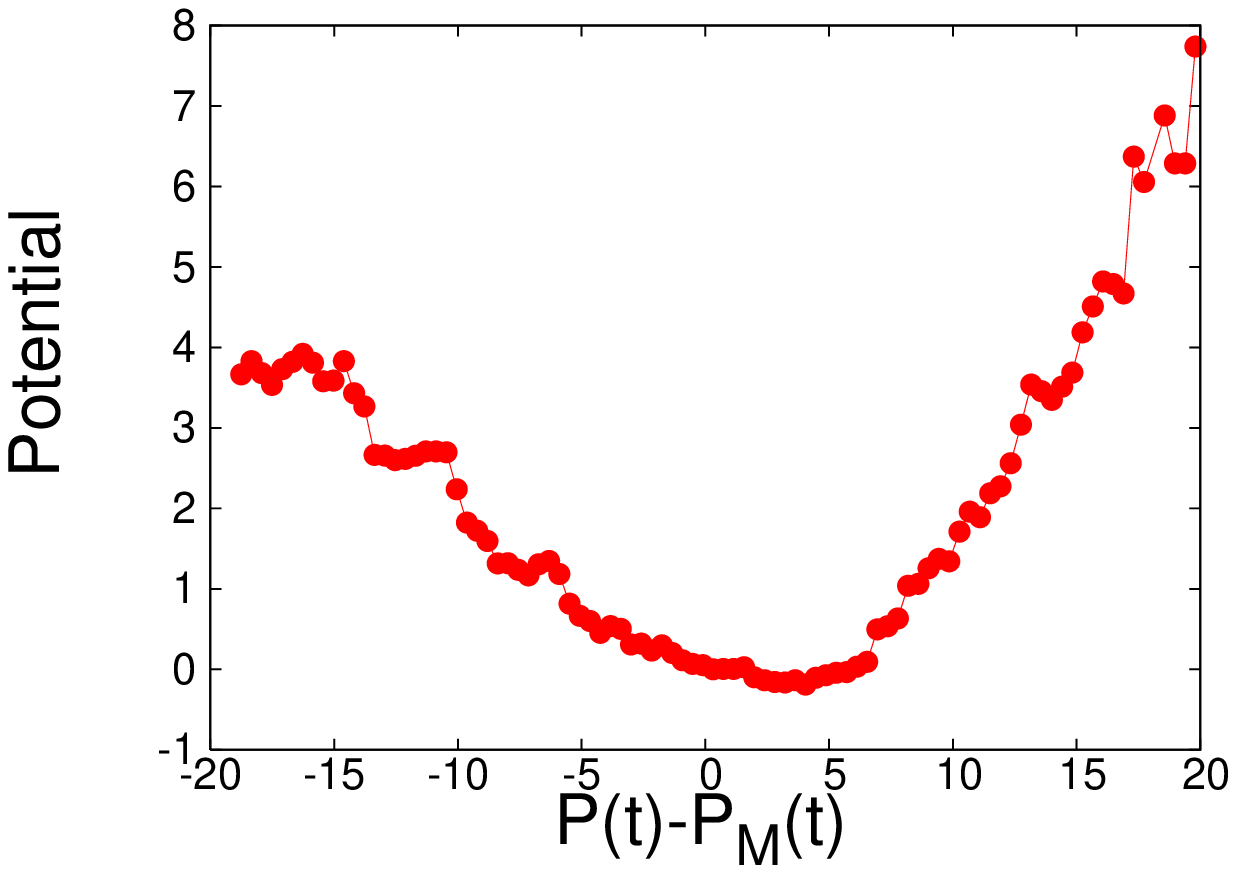}}
\subfigure[]{
\label{fig:maj}
\includegraphics[width=3in]{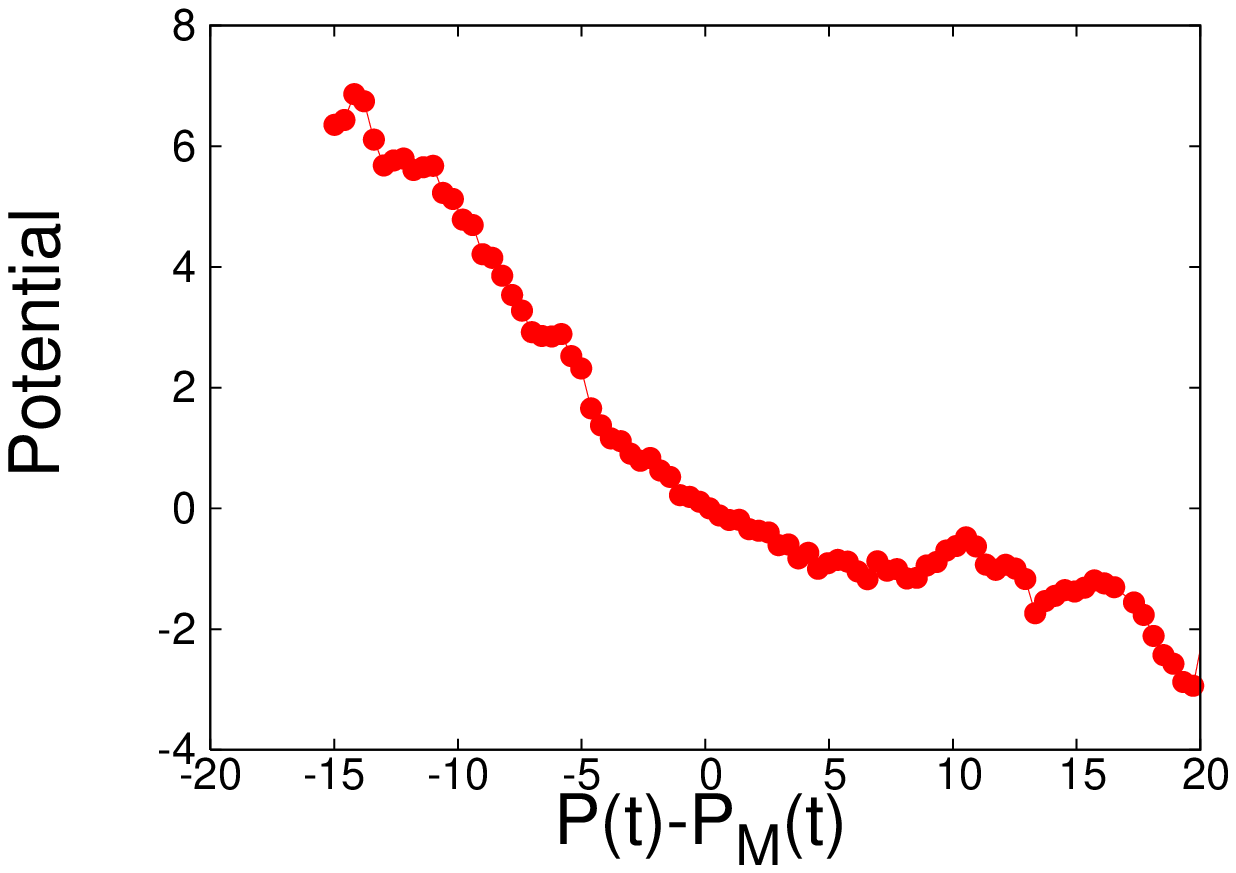}}
\caption{The potential analysis with $M=20$ for the two periods indicated in
  Fig.~\ref{fig:grafminmaj} is shown. The analysis for the minority
region leads to an attractive (even though not really quadratic)  potential.
Instead, analyzing the majority region we found a repulsive potential.}
\label{fig:minmaj}
\end{figure}

In Fig.~\ref{fig:minmaj} are plotted the  potentials obtained performing
the effective potential analysis with $M=20$.
We can observe that, when the market is dominated by contrarians,
we obtain an attractive potential.
This shape of the potential reproduce the agents' tendency to keep
the price near its ``fundamental'' value.
We can also note that this potential is not 
perfectly quadratic as in model described in Eqs.~\ref{eq:stoch}
and ~\ref{eq:potential}.
In fact, plotting different potentials with various values of $M$
we can not obtain a data collapse scaling the potentials
with the factor $(M-1)$.
In case of market dominated  by trend followers, we can observe the
presence of well defined trends (bubbles and crashes).
In this case the agents try to follow the trends and the price
tends to go away from his fundamental value.
In this case we obtain a repulsive potential.

Therefore, the potential analysis is able to detect the
agents' behavior based on microscopic rules only analyzing the data of 
a macroscopic
variable, $P(t)$.

From the viewpoint of modeling real markets, it will be very
interesting to introduce an agent based model in which agents perform
their decision (buy/sell) by considering their expectations about the
next price increment  using  Eq.~\ref{A_rel}, with different constant
of proportionality and different values of the 'memory' M and  not on
the basis of a set of given strategies (as in the minority game
framework).
Work along these lines is currently in progress.

\section{Results for Real Stock Prices from NYSE}

\begin{figure}[h]
\centering
\subfigure[]{
\label{fig:realpoteplotA}
\includegraphics[width=2in]{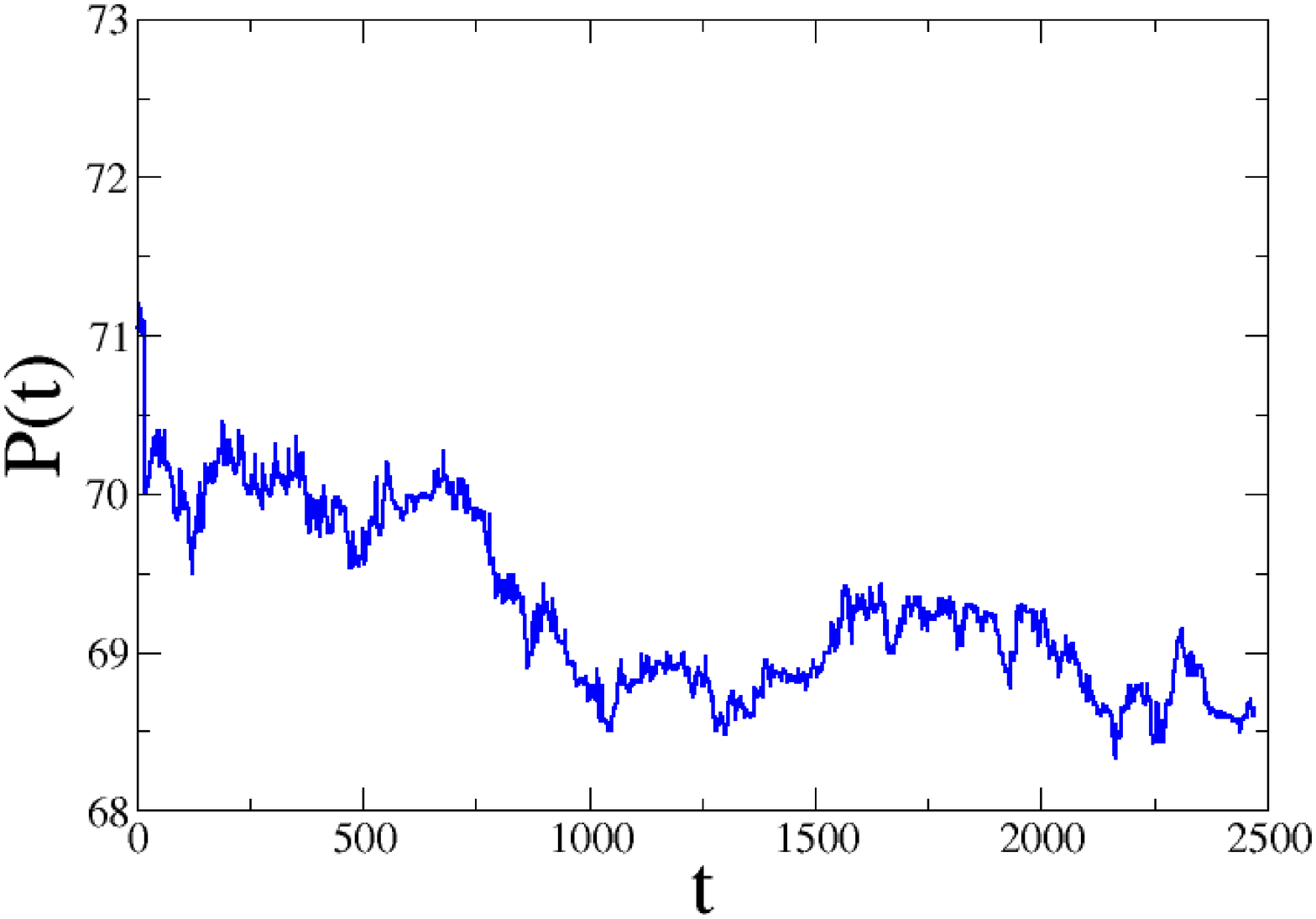}}
\subfigure[]{
\label{fig:realpoteplotB}
\includegraphics[width=2in]{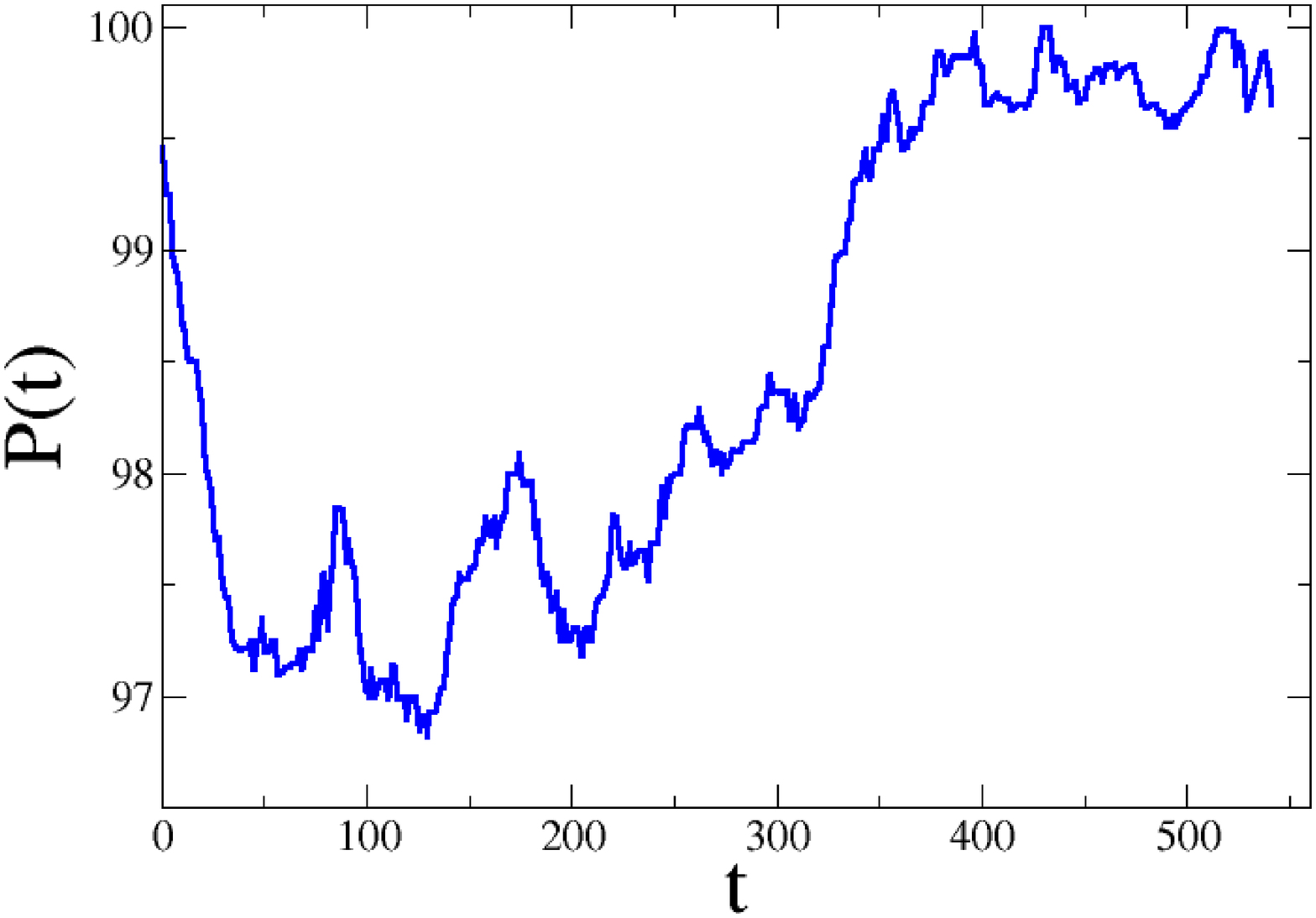}}
\subfigure[]{
\label{fig:realpoteplotC}
\includegraphics[width=2in]{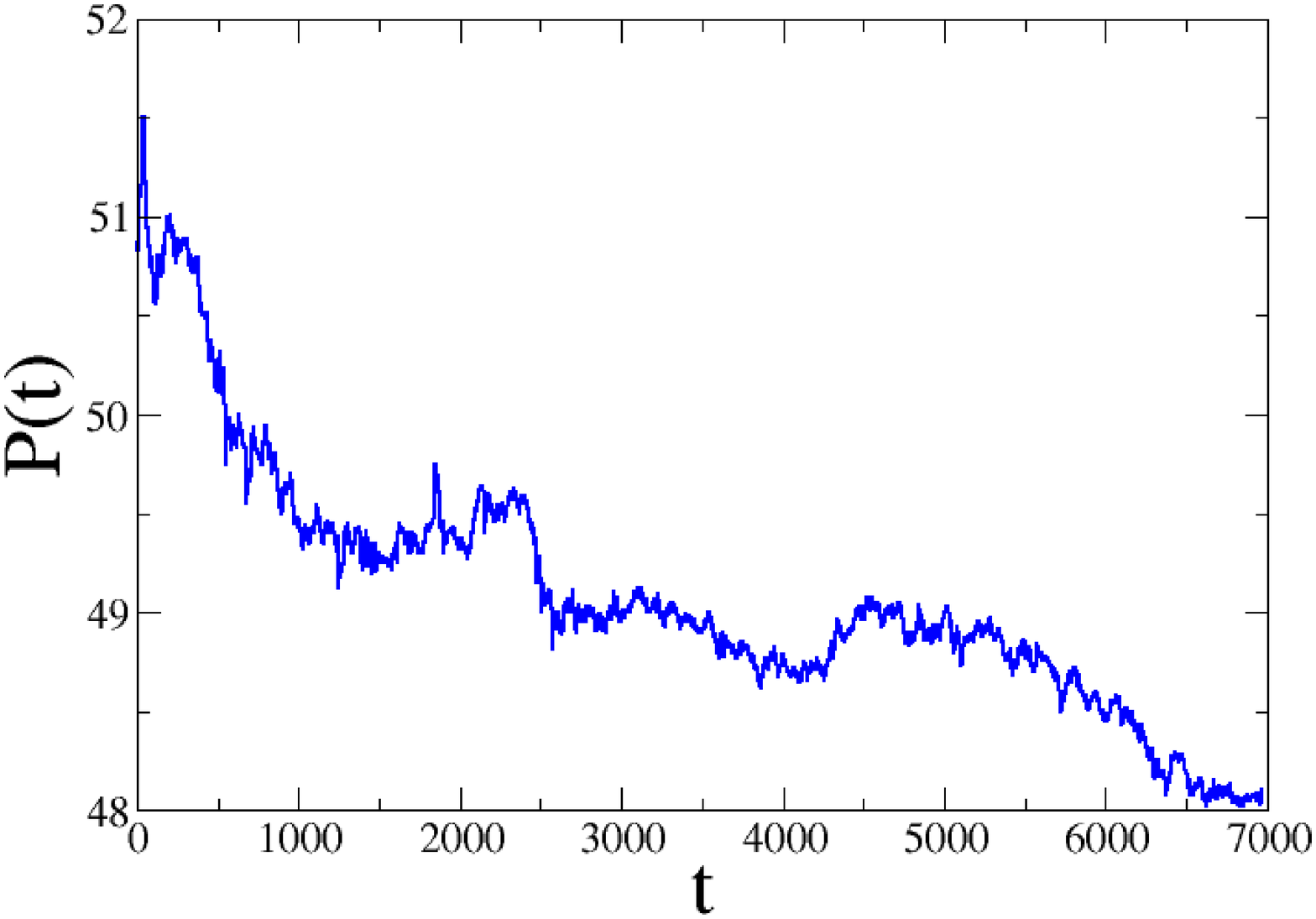}}
\caption{The time evolution of three stock index (CLF, ITU and TOL) is shown.
The time is expressed in tick and correspond to one trading day.}
\label{fig:realpoteplot}
\end{figure}

For our potential analysis we consider as database the price time series of 
all the transactions of a 
selection of 20 NYSE stocks.
These have been selected to be representative and with intermediate
volatility.
This corresponds to  volumes of $10^5-10^6$
stocks exchanged per day.
We consider 80 days from October 2004 to February 2005.

\begin{figure}[h]
\centering
\subfigure[]{
\label{fig:realpoteA}
\includegraphics[width=2in]{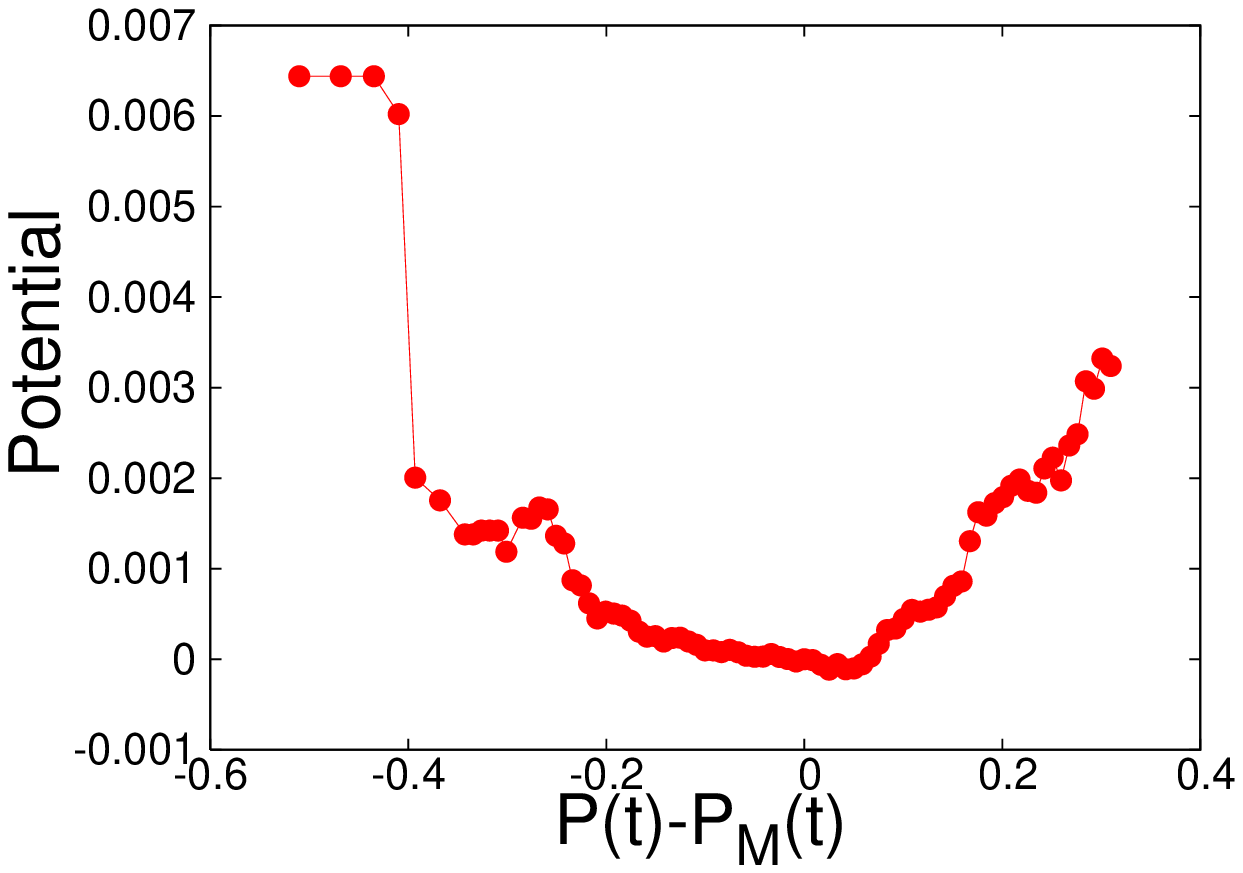}}
\subfigure[]{
\label{fig:realpoteB}
\includegraphics[width=2in]{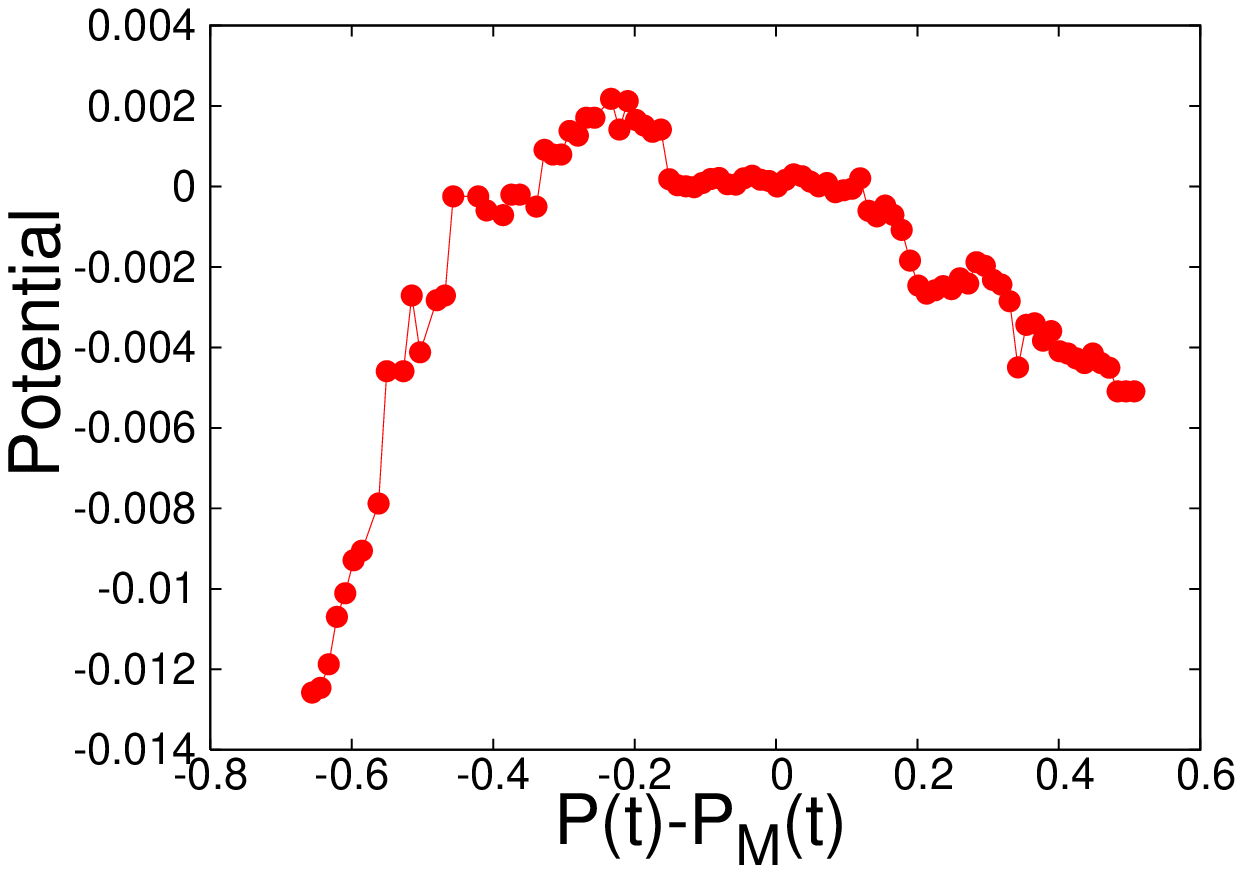}}
\subfigure[]{
\label{fig:realpoteC}
\includegraphics[width=2in]{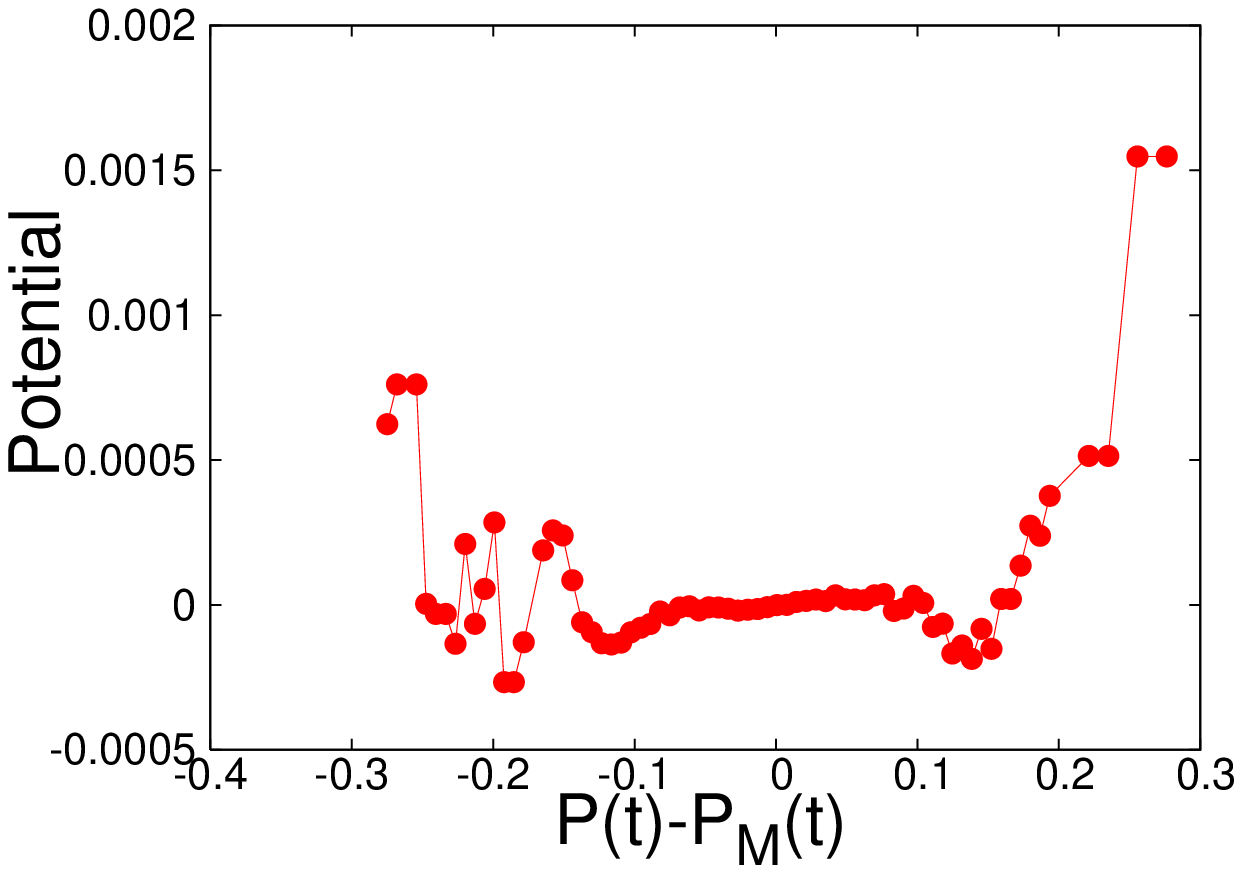}}
\caption{The effective potential method has been applied to the
price evolution of the three stocks plotted in Fig.~\ref{fig:realpoteplot}.
We can see that the effective potential are rather asymmetric and not all
quadratic. The potential in (a) seems quadratic and attractive in the central
part but has asymmetric tails. The potential in (b) and (c) are not quadratic.
In particular the potential in $b$ is asymmetric and looks like
piecewise linear as in Fig.~\ref{fig:signB}. The potential in (c) is flat
like a simple RW potential.} 
\label{fig:realpote}
\end{figure}

The time series we consider are by a sequential order tick by tick.
This is not identical to the price value as a function of
physical time but we have tested that the results
are rather insensitive to this choice.

The statistical properties of these kind of data are relatively
homogeneous within the time scale of a trading day but
the large jumps of the prices between different days prevent
the extension of the analysis to large times~\cite{vale2}.
So we focus our potential analysis considering the
stock-prices fluctuations within a trading day.
In Fig.~\ref{fig:realpoteplot} are plotted the time evolutions
of three stock indexes in a trading day.

If we perform the effective potential method for a trading day
of a given stock, we found shapes of the effective potential
that are very irregular and often asymmetric.
In Fig~\ref{fig:realpote} are plotted the results obtained for
the data plotted in Fig~\ref{fig:realpoteplot}.
We can see that the shapes of the potentials are not always quadratic.
The potential in Fig.~\ref{fig:realpoteA} it seems 
rather quadratic and attractive
while in  Fig.~\ref{fig:realpoteB} has a piecewise linear shape
similar to the potential plotted in Fig.~\ref{fig:signB}.
The potential in Fig~\ref{fig:realpoteC} seems flat as
one expects from a simple RW model.

\begin{figure}[h]
\centering
\subfigure[]{
\label{fig:mean1}
\includegraphics[width=3in]{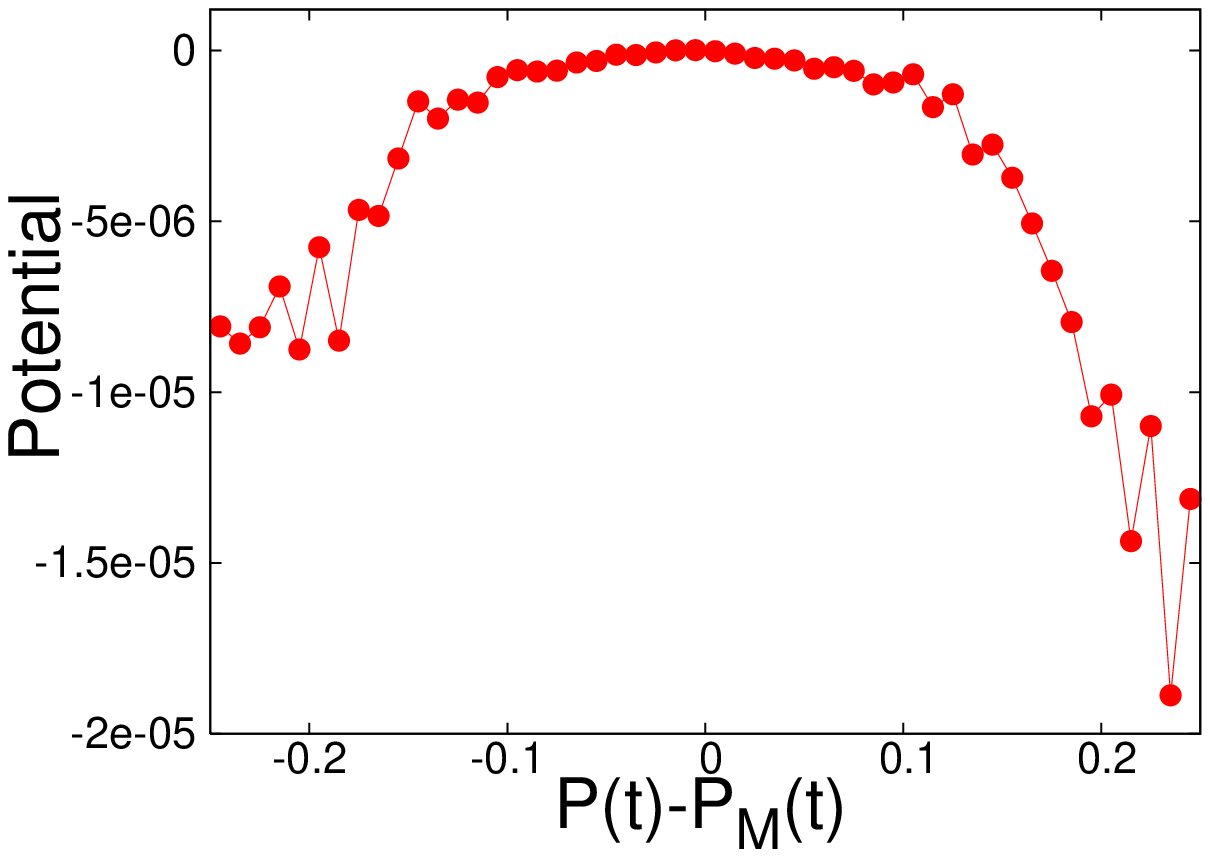}}
\subfigure[]{
\label{fig:mean2}
\includegraphics[width=3in]{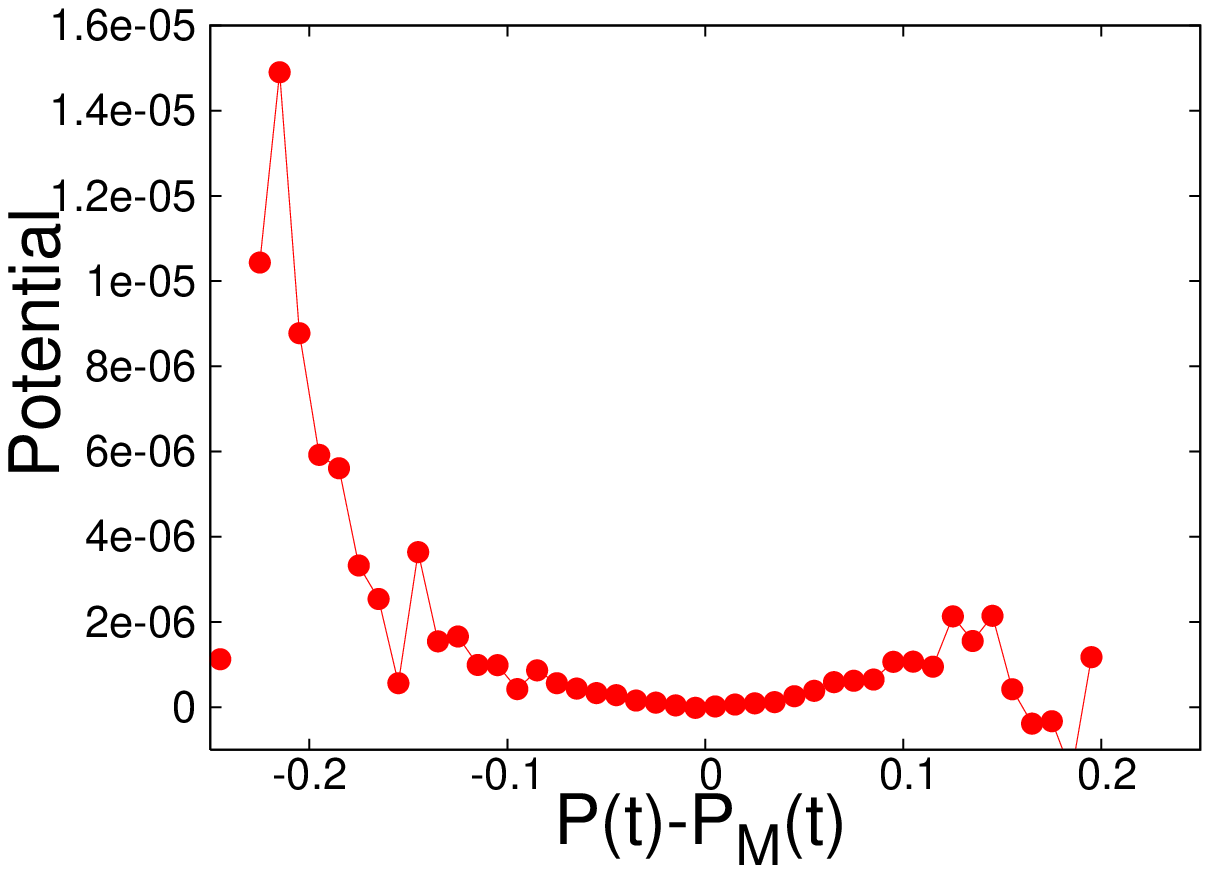}}
\caption{This plot shows the average shape of the potential
over 80 trading days for two different stock indexes (BRO and PG).
The shape is quite quadratic and symmetric, and in (a) is repulsive
while in (b) attractive.}
\label{fig:mean}
\end{figure}

Instead, if we consider some average over a long period (80 trading days)
of the potentials obtained for a single day, the resulting potentials seems
to be quadratic as in ~\cite{taka1}.
In Fig.~\ref{fig:mean} are shown the average shape of
the potential for two stock indexes (BRO and PG).
We can observe a rather quadratic shape for
the potential.
In Fig.~\ref{fig:mean1} the potential is quadratic and repulsive
while for the index PG the potential is attractive, as
we can see in Fig.~\ref{fig:mean2}.

\bibliographystyle{unsrt}

\bibliography{Alfi}

\end{document}